\documentstyle[twocolumn,floats,aps,prb,epsf]{revtex}		
\draft
\title{\boldmath $^{27}\!$Al Impurity-Satellite NMR and Non-Fermi-Liquid Behavior \\ 
in U$_{1-x}$Th$_x$Pd$_2$Al$_3$}
\author{Chia-Ying Liu and D.~E. MacLaughlin}
\address{Department of Physics, University of California, Riverside, California 92521}
\author{H.~G. Lukefahr, G. Miller, and K. Hui}
\address{Whittier College, Whittier, California 90608}
\author{M.~B. Maple, M.~C. de~Andrade, and E.~J. Freeman}
\address{University of California, San Diego, California 92093-0319}

\author{\small(January 7, 2000)}				
\address{\parbox{14cm}{\bigskip\rm\small				
Non-Fermi-liquid (NFL) behavior in the $f$-sublattice-diluted alloy system~U$_{1-x}$Th$_x$Pd$_2$Al$_3$ has been studied using $^{27}\!$Al nuclear magnetic resonance (NMR). Impurity satellites due to specific U near-neighbor configurations to $^{27}\!$Al sites are clearly resolved in both random and field-aligned powder samples. The particular configuration associated with each satellite is identified by comparison of calculated and observed satellite intensities. The spatial mean~$\overline{K}$ and rms spread~$\delta K$ of impurity satellite shifts, which are related to the mean~$\overline{\chi}$ and rms spread~$\delta\chi$ of the inhomogeneous susceptibility, have been measured in field-aligned powders with the crystalline $c$ axis both perpendicular and parallel to the external field. Satellites corresponding to only one uranium near neighbor were chosen for analysis, since in this case $\delta\chi(T)/\overline{\chi(T)} = \delta K/\overline{K}$ independent of the (unknown) spatial correlation of the random susceptibility inhomogeneity. The relatively narrow lines observed at low temperatures suggest that disorder-induced inhomogeneity of the $f$-ion--conduction-electron hybridization is not the cause of NFL behavior in these alloys: at low temperatures the experimental values of $\delta\chi(T)/\overline{\chi(T)}$ are much smaller than required by disorder-driven models. This is in contrast to results in at least some alloys with disordered non-$f$-ion nearest neighbors to $f$ ions (``ligand disorder''), where disorder-driven theories give good accounts of NFL behavior. Our results suggest that $f$-ion dilution does not produce as much inhomogeneity of the hybridization strength as substitution on ligand sites.
\\[6pt] PACS numbers: 71.27.+a, 75.30.Mb, 76.60.Cq.}}		
\begin{document}\maketitle						

\section{Introduction}

The magnetism of heavy-fermion $f$-electron materials is quenched at low temperatures by conduction-electron (Kondo) screening. A many-body ground state is formed that has traditionally been described within Landau's Fermi-liquid theory both for dilute alloys\cite{Nozi74} (the single-impurity Kondo problem) and for concentrated ``Kondo lattice'' systems (with or without lattice disorder). But the thermodynamic and transport properties of a number of $f$-electron ``heavy-fermion'' metals and alloys do not behave as predicted by Fermi-liquid theory.\cite{ITP96} The inapplicability of this picture is signaled by (a)~weak power-law or logarithmic divergences of the specific heat Sommerfeld coefficient~$\gamma(T) = C(T)/T$ and the magnetic susceptibility~$\chi(T)$, both of which are constant in Fermi-liquid theory, and (b)~a temperature dependence of the electrical resistivity which is weaker than the $T^2$ prediction of Fermi-liquid theory. Better theoretical and experimental understanding of these so-called non-Fermi-liquid (NFL) systems has been the goal of a considerable amount of research in recent years.

Two broad classes of theoretical explanation of NFL behavior have emerged: (a)~proximity to a zero-temperature quantum critical point (QCP), of either a single-ion or a cooperative nature,\cite{ITP96,TsRe93} and (b)~the effect of lattice disorder on the Kondo properties of the $f$ ions.\cite{ITP96,BMLA95,MBL96,MDK96,CNCJ98,dACDD98,AHCN00} In the QCP picture the NFL behavior is due to quantum fluctuations associated with a critical point at zero temperature. This mechanism is operative in uniform systems, since disorder is not required. In disorder-driven scenarios the effect of structural disorder on $f$-electron many-body effects such as the Kondo effect and the Ruderman-Kittel-Kasuya-Yosida (RKKY) interaction between $f$ ions produces a broad inhomogeneous distribution of the local susceptibilities~$\chi_j$ associated with the $f$ ions. Uncompensated Kondo ions far from the singlet ground state, which are not described by Fermi-liquid theory, give rise to large values of $\chi_j$ and the NFL properties of the material. QCP and disorder-driven mechanisms need not be mutually exclusive, however, since critical fluctuations of a disordered system might also be involved in NFL behavior. 

The single-ion ``Kondo disorder'' picture,\cite{BMLA95,MBL96,MDK96} in which the uncompensated ions are assumed not to interact, was developed to explain nuclear magnetic resonance (NMR) experiments in the NFL alloys~UCu$_{5-x}$Pd$_s$, $x = 1.0$ and 1.5, that revealed wide distributions of frequency shifts reflecting the required susceptibility inhomogeneity. In the Kondo disorder model structural disorder gives rise to a wide distribution of local Kondo temperatures~$(T_K)_j$; this leads to a distribution of $\chi_j$ which becomes correspondingly wide at low temperatures. The ``Griffiths-phase'' model of Castro~Neto {\em et al.\/}\cite{CNCJ98,dACDD98,AHCN00} takes into account RKKY interactions between uncompensated $f$-ion moments in the disordered system. These RKKY interactions couple the uncompensated moments into clusters, the thermal behavior of which can be described in terms of Griffiths singularities\cite{Grif69} associated with the distribution of cluster sizes. 

Both the Kondo-disorder and Griffiths-phase theories make definite predictions of the inhomogeneous spread in magnetic susceptibility that is capable of being measured by magnetic resonance techniques such as NMR\@. The present paper compares these predictions with $^{27}\!$Al NMR spectra in unaligned and field-aligned powder samples of the NFL alloys~U$_{1-x}$Th$_x$Pd$_2$Al$_3$, $x = 0.7$, 0.8, and 0.9. 

\begin{figure}% 
%[p] \epsfxsize 6.5in \epsfbox{Liufig1.eps}			% for preprint version
[t] \epsfxsize \columnwidth \epsfbox{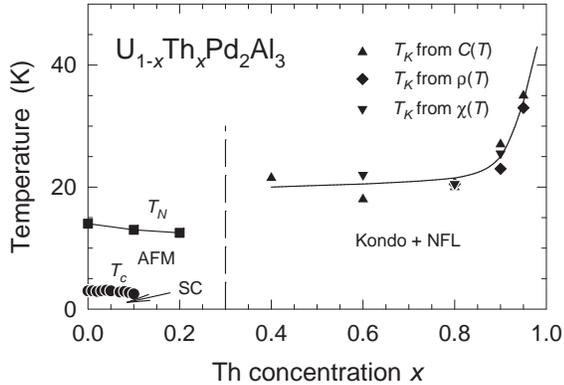}		% for two-column version
\caption{Phase diagram of U$_{1-x}$Th$_x$Pd$_2$Al$_3$. Antiferromagnetic (AFM), Superconducting (SC), and Kondo/non-Fermi-\protect\linebreak[2] liquid (Kondo + NFL) regions are shown. NFL behavior occurs for $x \protect\gtrsim 0.6$, where the Kondo temperature~$T_K$ was determined from measurements of specific heat~$C(T)$, electrical resistivity~$\rho(T)$, and magnetic susceptibility~$\chi(T)$. Data from Ref.~\protect\onlinecite{MdAHD95}.}\label{fig:phasediag}
\end{figure}

The isostructural alloy series U$_{1-x}$Th$_x$Pd$_2$Al$_3$ exhibits NFL behavior for intermediate to high thorium concentrations.\cite{MdAHD95} The phase diagram of U$_{1-x}$Th$_x$Pd$_2$Al$_3$ is shown in Fig.~\ref{fig:phasediag}. 
The heavy-fermion end compound UPd$_2$Al$_3$ exhibits coexistence of antiferromagnetic order ($T_N = 14$ K) and superconductivity\cite{GSTK91} ($T_c = 2$ K) that persists for low Th concentrations ($x < 0.2$). For $x > 0.6$ the electrical resistivity, specific heat, and magnetic susceptibility are all indicative of NFL behavior.\cite{DaMa94}

The RKKY coupling between $^{27}\!$Al nuclei and U-ion spins affects the $^{27}\!$Al NMR in a number of ways, of which the most important for our purposes is the paramagnetic shift~$K$ of the field for resonance at fixed frequency;\cite{CBK77,KSdef} this shift is expected to be proportional to the U-ion susceptibility~$\chi$. Any inhomogeneity in $\chi$ leads to a corresponding distribution of shifts that broadens the NMR line. The relation between the rms spread~$\delta\chi$ in $\chi$ and the NMR linewidth~$\sigma$ is described elsewhere,\cite{MBL96,BMAF96,LMCN00} where it is shown that $\sigma/(\overline{K}H_0)$, where $\overline{K}$ is the spatially averaged relative shift and $H_0$ is the applied field, is an estimator for the fractional rms spread~$\delta\chi/\overline{\chi}$, where $\overline{\chi}$ is the spatially averaged susceptibility. We can write $\sigma/(\overline{K}H_0)$ equivalently as $\delta K/\overline{K}$, where $\delta K \equiv \sigma/H_0$ is the relative rms spread in shifts. 

NMR line broadening can also arise from dynamic (lifetime) effects due to nuclear spin-lattice relaxation. Pulsed NMR techniques can be used to estimate such lifetime broadening independently of the spectral width,\cite{CBK77} and we have found that in U$_{1-x}$Th$_x$Pd$_2$Al$_3$ spin-lattice relaxation is far too small to contribute significantly to the observed spectral linewidths.

In a diluted magnetic alloy such as U$_{1-x}$Th$_x$Pd$_2$Al$_3$ the NMR shift is also distributed because of the spatial dependence of the RKKY interaction and the random positions of the magnetic ions, even if the susceptibility associated with these ions were uniform. Thus additional broadening due to susceptibility inhomogeneity may be difficult to resolve. This is unlike the situation in an alloy with {\em ligand disorder\/} (i.e., disorder in the nonmagnetic ions of the compound), where the magnetic-ion sublattice is ordered. In this case the only source of broadening is susceptibility inhomogeneity, as long as the ligand disorder does not affect the RKKY interaction significantly. Even if it does the two broadening mechanisms can be distinguished.\cite{LMCN00}

If, however, specific near-neighbor magnetic-ion configurations of the observed nuclei are probable enough in a dilute alloy, and if the shifted NMR frequencies of these nuclei are large enough to be resolved, rather than merely contributing to ``dilution broadening'' of the resonance line, then the shifts and linewidths of these {\em impurity satellites\/} may be studied separately.\cite{BoSl76} Impurity satellites provide a much better characterization of the inhomogeneous susceptibility distribution in an alloy with $f$-ion sublattice dilution (such as U$_{1-x}$Th$_{x}$Pd$_2$Al$_3$) than is possible if the satellites are not resolved. 

An important result of the present studies was the observation of resolved $^{27}\!$Al impurity satellites in U$_{1-x}$Th$_x$Pd$_2$Al$_3$. The shifts and linewidths of these satellites have been analyzed to provide information on the inhomogeneous distribution of local susceptibility in this system. We find that the width of the susceptibility distribution is considerably smaller than required to explain NFL behavior, and a different mechanism must be sought. This result suggests that, in contrast to the situation in ligand-disorder alloys, $f$-sublattice disorder may not produce enough inhomogeneity in the $f$-electron--conduction-electron hybridization to drive the NFL behavior.

\section{Disorder-driven NFL theories} \label{sect:disorderNFL}

In the following we briefly describe the single-ion Kondo-disorder\cite{BMLA95,MBL96,MDK96} and Griffiths-phase\cite{CNCJ98,dACDD98,AHCN00} models, after which we consider their applicability to the NFL behavior in U$_{1-x}$Th$_{x}$Pd$_2$Al$_3$.

The simplest implementation of the single-ion Kondo disorder model assumes that the $f$ ions are coupled to the conduction-electron bath by a random distribution of Kondo coupling constants $g = \rho{\cal J}$, where $\rho$ is the conduction-electron density of states at the Fermi energy and $\cal J$ is the local-moment--conduction-electron exchange energy. Then $g$ is related to the Kondo temperature~$T_K$ by
\begin{equation} T_K = E_F\,\exp(-1/g) \,, \label{eq:T_K} \end{equation}
where $E_F$ is the Fermi energy of the host metal. Thus a modest distribution of $g$ can give rise to a broad distribution of $T_K$ if most values of $g$ are small. If the distribution function $P(T_K)$ is broad enough so that $P(T_K{= }0)$ does not vanish, then at any nonzero temperature $T$ those $f$ ions for which $T_K < T$ will not be compensated. Fermi-liquid theory does not apply to them, and they give rise to the NFL behavior. 

Such uncompensated $f$ ions dominate thermal and transport properties at low temperatures. The magnetic susceptibility is correspondingly distributed, as can be seen from the Curie-Weiss law
\begin{equation} \chi(T,T_K) = \frac{\cal C}{T + \alpha T_K} \end{equation}
($\alpha \sim 1$) that approximately characterizes the Kondo physics, and is inhomogeneous on an atomic scale. Fitting this model to the temperature dependence of the bulk (i.e., spatially averaged) magnetic susceptibility~$\overline{\chi(T)}$ gives the distribution of $g$, which is then used to predict $\delta\chi(T)/\overline{\chi(T)}$.\cite{BMLA95,MBL96}

In the Griffiths-phase theory various physical properties are predicted to diverge at low temperatures as weak power laws of temperature. For example, the electronic specific heat~$C(T)$ and the spatially averaged magnetic susceptibility~$\overline{\chi(T)}$ are given by
\begin{equation}
C(T)/T \propto \overline{\chi(T)}\propto T^{-1+\lambda } \,, \label{eq:Grifchi} 
\end{equation}
and the fractional rms spread~$\delta\chi(T)/\overline{\chi(T)}$ is given by
\begin{equation}
\frac{\delta\chi(T)}{\overline{\chi(T)}} \propto T^{-\lambda /2}\,. \label{eq:GrifdXonX}
\end{equation}
The nonuniversal exponent~$\lambda$ is a parameter that determines the degree of NFL character. The Griffiths phase is characterized by $\lambda \le 1$, so that the susceptibility diverges at zero temperature (the case~$\lambda = 1$ is marginal and gives rise to logarithmic divergences). Since $C(T)/T$ and $\overline{\chi(T)}$ have the same temperature dependence the Wilson ratio~$\overline{\chi(T)}T/C(T)$ is independent of temperature in this picture. The procedure used to compare the Griffiths-phase theory with experiment is similar to that described above for the Kondo-disorder analysis: bulk susceptibility data are fit to the corresponding theoretical expressions, and the calculated $\delta\chi(T)/\overline{\chi(T)}$ is compared with NMR data.

\section{Experiment}

\begin{figure}%
%[p] \epsfxsize 6.5in \epsfbox{Liufig2.eps}			% for preprint version
[t] \epsfxsize \columnwidth \epsfbox{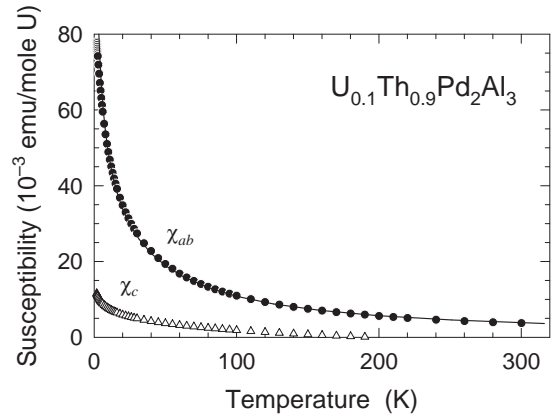}		% for two-column version
\caption{Temperature dependence of the anisotropic susceptibility in U$_{0.1}$Th$_{0.9}$Pd$_2$Al$_3$. Circles: $ab$-plane susceptibility $\chi_{ab} (T)$. Triangles: $c$-axis susceptibility $\chi_c(T)$. Curve: fit of Griffiths-phase model (Refs.~\protect\onlinecite{CNCJ98,dACDD98,AHCN00}) to $\chi_{ab} (T)$.}
\label{fig:suscept}
\end{figure}

Samples of U$_{1-x}$Th$_{x}$Pd$_2$Al$_3$, $x = 0.7$, 0.8, and 0.9, were prepared as described previously.\cite{MdAHD95,DaMa94} The arc-melted ingots were crushed and passed through a 100-$\mu$m sieve. $^{27}\!$Al NMR experiments were carried out on unaligned powders and also on epoxy-cast powder samples in which the single-crystal powder grains were aligned by a 6-T magnetic field during hardening of the epoxy. The crystal symmetry is hexagonal (space group~$P6/mmm$) and thus the susceptibility is uniaxially anisotropic. The direction of largest magnetic susceptibility is in the basal ($ab$) plane, so that in order to orient the $c$ axes of the grains it was necessary to rotate the sample around an axis perpendicular to the magnetic field while the epoxy hardened.\cite{LMSH93} The rotation axis then defines the $c$-axis orientation of the grains. The anisotropic susceptibility obtained from a field-aligned sample of U$_{0.1}$Th$_{0.9}$Pd$_2$Al$_3$ is shown in Fig.~\ref{fig:suscept}, where the strong anisotropy that permits the field alignment can be seen.

Also shown in Fig.~\ref{fig:suscept} is the result of fitting the Griffiths-phase model prediction for the uniform susceptibility to the experimental $\chi_{ab} (T)$ data, as described above in Sec.~\ref{sect:disorderNFL}. Aside from an overall scale factor the fit parameters are the exponent~$\lambda$ and a high-energy cutoff~$\epsilon_0$ for the distribution of cluster energies. Best fit was found for $\lambda = 0.90$ and $\epsilon_0/k_B = 130$ K\@. A good fit was also obtained with the Kondo-disorder model, where the parameters describing the disorder were taken to be the Fermi energy~$E_F$ [cf.\ Eq.~(\ref{eq:T_K})] and the mean~$\overline{g}$ and standard deviation~$\delta g$ of an assumed Gaussian distribution of coupling constants. The values~$E_F = 1$ eV, $\overline{g} = 0.149$, and $\delta g = 0.020$ gave the best fit (not shown). The results of these fits were used to calculate numerical values of $\delta\chi(T)/\overline{\chi(T)}$, which are compared with results of the NMR experiments in Sec.~\ref{sect:disorder}.

Field-swept $^{27}\!$Al NMR spectra were obtained using pulsed-NMR spin-echo signals and the frequency-shifted and summed Fourier transform processing technique described by Clark {\em et al.}\cite{CHLS95} The echo-decay lifetime~$T_2$ was found to be sufficiently long (hundreds of $\mu$s) so that no correction for the echo decay was needed. 

\subsection{\boldmath Spurious phases or impurity satellites in U$_{1-x}$Th$_x$Pd$_2$Al$_3$?}

\begin{figure}%
%[p] \epsfxsize 6.5in \epsfbox{Liufig3.eps}			% for preprint version
[t] \epsfxsize \columnwidth \epsfbox{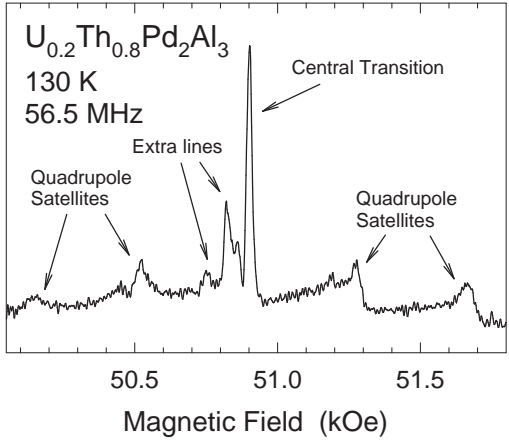}		% for two-column version
\caption{Representative $^{27}\!$Al quadrupole-split NMR spectrum from an unaligned powder sample of U$_{0.2}$Th$_{0.8}$Pd$_2$Al$_3$. The spectrum is characterized by a narrow central transition and broad powder-pattern quadrupole satellites (Ref.~\protect\onlinecite{CBK77}). Extra lines are visible on the low-field side of the central transition, and to some extent on the low-field sides of the quadrupole satellites.}
\label{fig:ppspect}
\end{figure}

Figure~\ref{fig:ppspect} shows as an example a spectrum from an unaligned powder sample of U$_{0.2}$Th$_{0.8}$Pd$_2$Al$_3$. 
In U$_{1-x}$Th$_{x}$Pd$_2$Al$_3$ the $mmm$ point symmetry of the Al site is lower than cubic, so that the $^{27}\!$Al nuclear Zeeman levels are split by the quadrupole interaction between the $^{27}\!$Al nuclear quadrupole moment~$Q$ and the crystalline electric field gradient~$q$.\cite{CBK77} Then quadrupole satellite resonances, corresponding to $m_I \leftrightarrow m_I{-}1$ ($m_I \ne 1/2$) nuclear spin transitions, are observed in the form of broad peaks shifted from a narrow central $(1/2 \leftrightarrow -1/2)$ transition. The positions of the quadrupole satellites are shifted to first order in the coupling constant~$e^2qQ$ and depend on crystallite orientation, so that the quadrupolar satellites are ``powder-pattern broadened'' by the random orientations of the powder grains in the sample. The central transition is shifted only to second order in $e^2qQ$, however, and suffers much less powder-pattern broadening than the satellites when $e^2qQ$ is smaller than the $^{27}\!$Al nuclear Zeeman frequency. 

The most important feature of Fig.~\ref{fig:ppspect} is the group of extra lines on the low-field side of the central transition. (There is also a hint of this structure in some of the quadrupole satellites.) The extra lines, which also appear for Th concentrations $x = 0.7$ and $0.9$, were initially suspected to be due to spurious metallurgical phases, although x-ray powder diffraction measurements indicated the samples were single-phase.\cite{DaMa94} More than one extra ``minority'' line was observed, but in the following only the most intense minority line is compared with the ``principal'' central transition (Fig.~\ref{fig:ppspect}). 

\begin{figure}%
%[p] \epsfxsize 6.5in \epsfbox{Liufig4.eps}				% for preprint version
[t] \epsfxsize \columnwidth \epsfbox{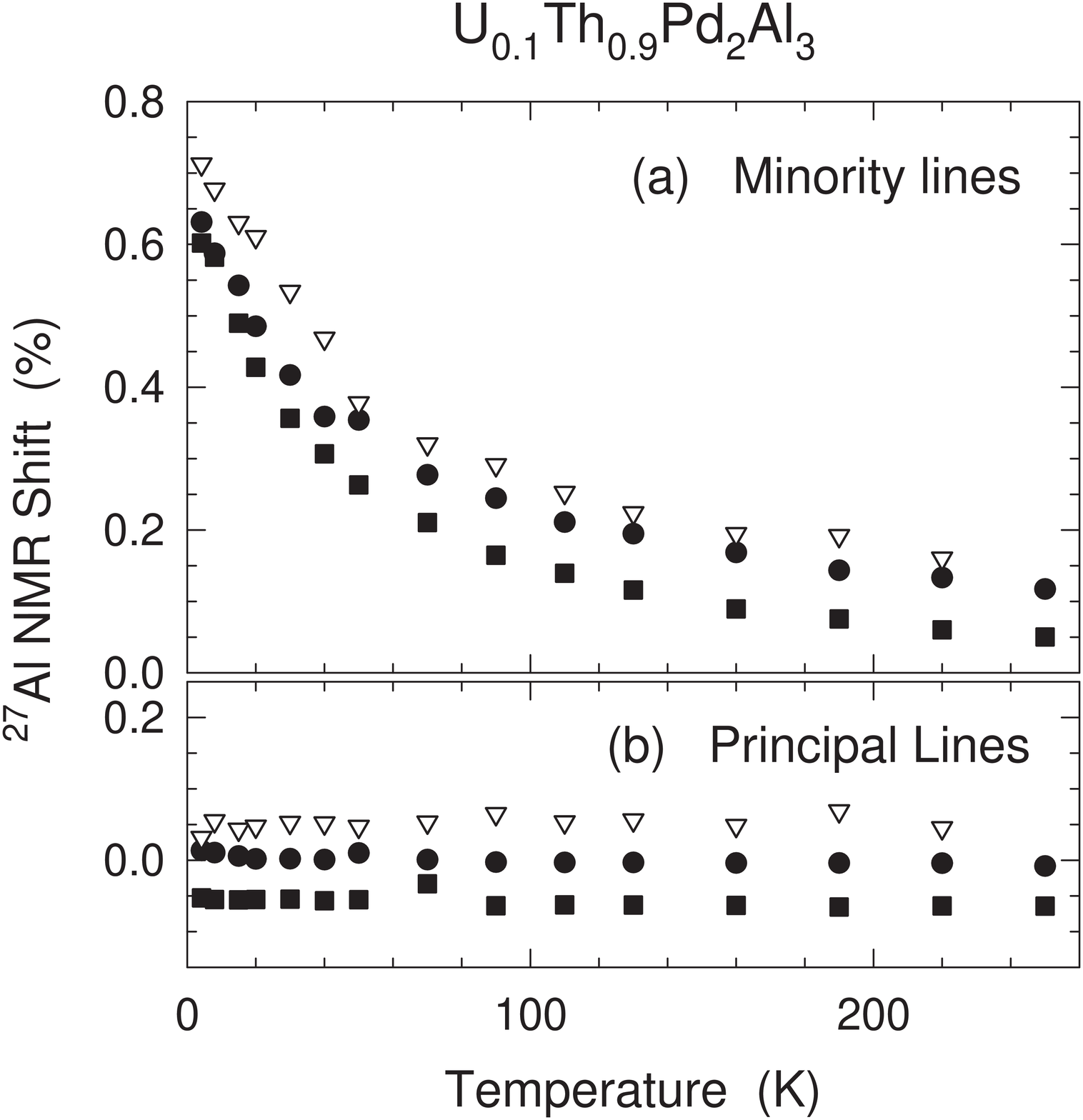}			% for two-column version
\caption{Temperature dependence of $^{27}\!$Al NMR shifts in unaligned powder samples of U$_{1-x}$Th$_x$Pd$_2$Al$_3$, $x = 0.7$ (squares), 0.8 (triangles), and 0.9 (circles). (a) Minority lines. (b) Principal lines.}
\label{fig:ppKS}
\end{figure}

The temperature dependence of the principal and minority shifts is shown in Fig.~\ref{fig:ppKS}. 
It can be seen that the shift of the principal line is independent of temperature, whereas the minority line exhibits a Curie-Weiss-like temperature-dependent shift. If the minority line were due to a spurious phase, then these results suggest that the principal phase is a nearly pure thorium compound, whereas the spurious phase or phases have a high uranium concentration. This seems rather unlikely, given the absence of x-ray evidence for phase segregation and the chemical similarity of thorium and uranium, and we take the results shown in Fig.~\ref{fig:ppKS} as initial evidence against the spurious-phase hypothesis. 

\begin{figure}%
%[p] \epsfxsize 6.5in \epsfbox{Liufig5.eps}			% for preprint version
[t] \epsfxsize \columnwidth \epsfbox{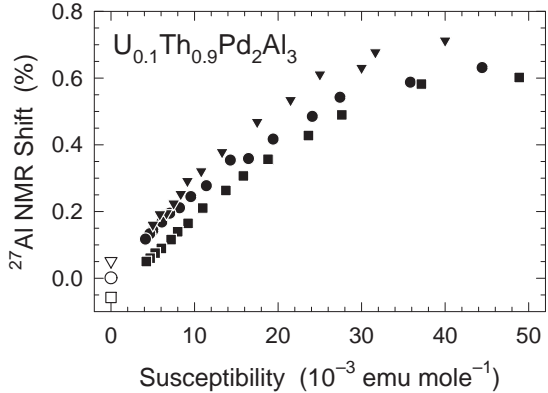}		% for two-column version
\caption{Clogston-Jaccarino plot of $^{27}\!$Al principal- and minority-line NMR shifts vs.\ bulk (spatially averaged) susceptibility in unaligned powder samples of U$_{1-x}$Th$_x$Pd$_2$Al$_3$, $x = 0.7$ (squares), 0.8 (triangles), and 0.9 (circles). Temperature is an implicit parameter. The minority-line data (filled symbols) show similar behavior for different Th concentrations~$x$. Extrapolations to zero susceptibility (high temperature) agree with the temperature-independent principal shifts (open symbols at zero susceptibility).}
\label{fig:ppClog}
\end{figure}

Figure~\ref{fig:ppClog} gives a Clogston-Jaccarino plot of shift versus susceptibility per mole uranium, with temperature an implicit variable. 
A linear relation between shift and $\overline{\chi(T)}$ is expected for impurity satellites if the transferred hyperfine field that couples the nuclear spin to the U moment is temperature-independent, and is indeed observed for small to moderate values of $\overline{\chi(T)}$. The slopes of these linear relations are more or less independent of thorium concentration~$x$, as expected for a local hyperfine interaction. [The nonlinear behavior for larger $\overline{\chi(T)}$ is not well understood, but may arise from changes in crystal-field level populations with temperature.]

The temperature independence of the principal-line shift is explained by the large number of configurations of more distant U ions and the oscillatory dependence of the RKKY interaction, which lead to broadening but little average shift contribution to the principal line.\cite{WaWa74} Varying the Th concentration~$x$ can change the host-metal band structure and lead to an $x$-dependent host Knight shift. The fact that the principal-line shifts do not vary monotonically with Th concentration is not well understood, but may be due to shift anisotropy together with sample-dependent preferential orientation of the powder grains.

Comparison of the principal- and minority-line shifts yields strong additional evidence that the minority lines are impurity satellites rather than due to spurious phases. One of the most striking features of Fig.~\ref{fig:ppClog} is the fact that extrapolations of the minority-line shifts to zero $\overline{\chi(T)}$ (infinite temperature) are in agreement with the temperature-independent principal-line shifts. This agreement would be a complete coincidence if the minority lines were due to spurious phases, but follows naturally if they are impurity satellites.

We conclude that the observed properties of the spectra establish the minority lines as impurity satellites rather than due to spurious phases. 

\subsection{Spectra from field-aligned samples}
The $^{27}\!$Al impurity-satellite spectra from unaligned powder samples of U$_{1-x}$Th$_x$Pd$_2$Al$_3$ described above are difficult to interpret, due to possible preferential orientation and broadening from shift anisotropy. We therefore used field-aligned samples for further NMR experiments. 

\begin{figure}%
%[p] \epsfxsize 6.5in \epsfbox{Liufig6.eps}			% for preprint version
[t] \epsfxsize \columnwidth \epsfbox{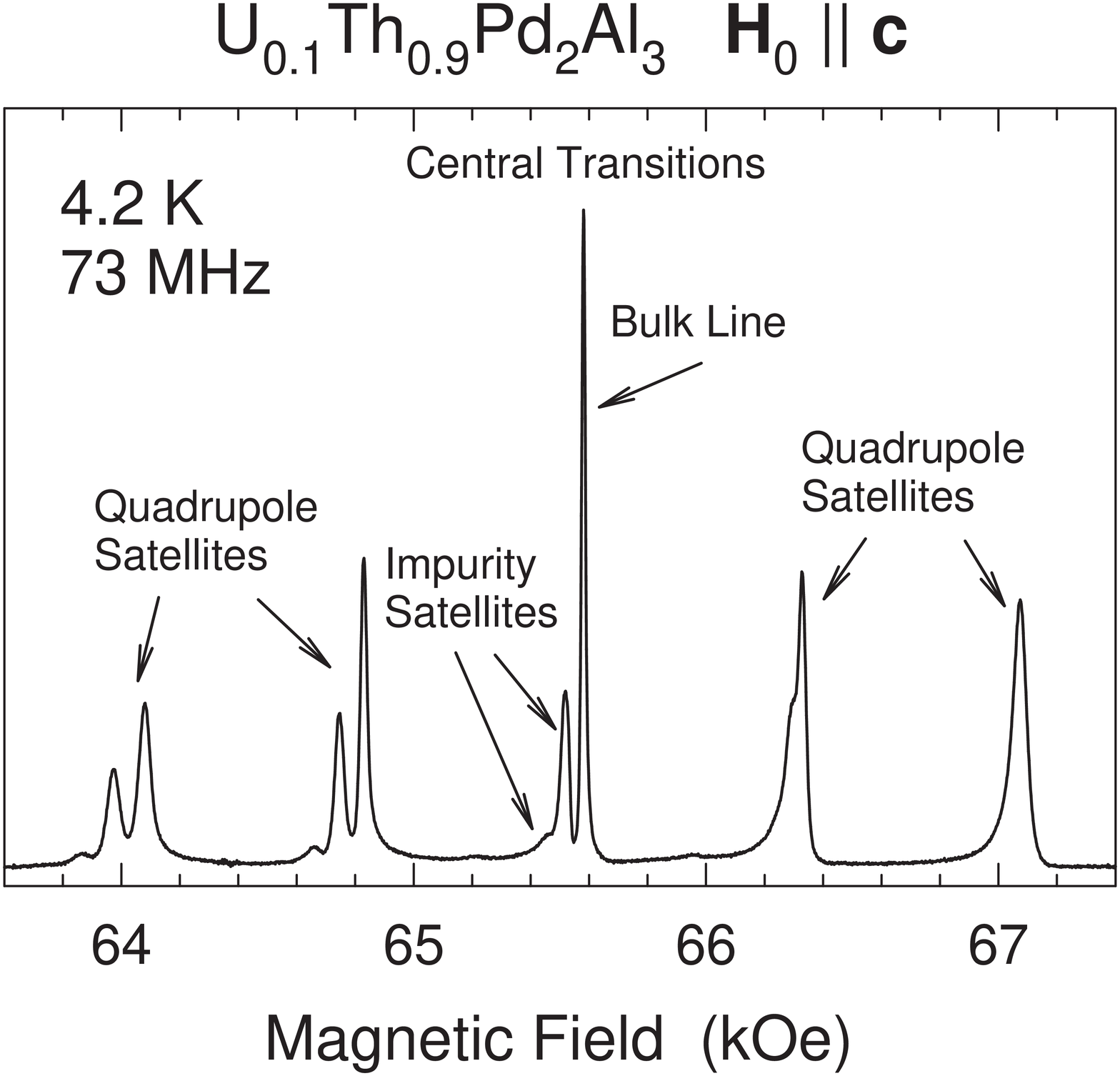}		% for two-column version
\caption{Representative $^{27}\!$Al NMR field-aligned spectrum from a field-aligned powder sample of U$_{0.1}$Th$_{0.9}$Pd$_2$Al$_3$ for ${\bf H}_0\parallel {\bf c}$. The bulk NMR lines and impurity satellites (indicated for the central transition only) are grouped into a central transition and four quadrupole satellites. The impurity satellites have slightly different anisotropic NMR shifts and quadrupole splittings than the bulk lines. The sharp lines indicate good alignment of the powder-grain $c$ axes.}
\label{fig:parrspect}
\end{figure}

Figure~\ref{fig:parrspect} shows a representative field-aligned spectrum from U$_{0.1}$Th$_{0.9}$Pd$_2$Al$_3$ for applied field~${\bf H}_0$ parallel to the $c$ axis (${\bf H}_0 \parallel {\bf c}$).
The sharp lines indicate good alignment of the powder-grain $c$ axes, and the impurity satellites are more apparent than in the unaligned powder spectra. The lines in the center of the spectrum are the $(1/2 \leftrightarrow -1/2)$ central transition bulk line and impurity satellites. Impurity satellites can also be seen associated with each quadrupole satellite. These are well resolved on the low-field side of the quadrupole-split spectrum but not on the high-field side, due to the combination of an anisotropic NMR shift and slightly larger quadrupole splittings (at fixed frequency) of the impurity satellites relative to the bulk lines. 

It can also be seen in Fig.~\ref{fig:parrspect} that the quadrupole satellites are somewhat broader than the central transition, due presumably to incomplete alignment and/or disorder in the quadrupole interaction; as mentioned above this results in first-order quadrupole broadening of the quadrupole satellites but only second-order broadening of the central transitions. Henceforth we consider only the central-transition bulk line and impurity satellites. 

Field-swept spectra were taken with a field range of about 300 Oe around the central transition and a small field step ($\sim$3 Oe) to have good resolution of narrow lines. Spectra were obtained as a function of temperature and field for both field directions (${\bf H}_0\parallel {\bf c}$ and ${\bf H}_0\perp {\bf c}$) on field-aligned samples of U$_{1-x}$Th$_x$Pd$_2$Al$_3$, $x = 0.7,0.8,$ and $0.9$. 

Impurity satellites and line shapes in dilute magnetic alloys have been treated by Walstedt and Walker,\cite{WaWa74} who showed on very general grounds that in the absence of susceptibility inhomogeneity the line shapes and widths of the bulk line and impurity satellites are the same. Inhomogeneity results in an increase of the satellite linewidths relative to that of the bulk line. We find that a Lorentzian line shape fits the bulk line best. Following the result of Walstedt and Walker, the impurity satellites were fit with a Lorentzian with the same width as the bulk line, convoluted with an extra (Gaussian) broadening function that describes the susceptibility inhomogeneity. 

Simple statistical considerations were used to constrain the fits. The probability~$P_n^{n_0} (x)$ of finding $n$ and only $n$ uranium impurities in a given near-neighbor (Th,U) shell around an Al site is given by 
\begin{equation}
P_n^{n_0} (x) = y^n(1-y)^{n_0-n}C_n^{n_0}\,, \label{eq:P(x)}
\end{equation}
where $y = 1-x$ is the U concentration, $n_0$ is the total number of U sites in the shell, and $C_n^{n_0} = n_0!/n!(n_0-n)!$ is the binomial coefficient. The intensity of each line (i.e., the area under the line) including the bulk line (for which $n = 0$ for all resolved near-neighbor shells) is proportional to this probability. The fitting procedure then consists of taking the number of most probable U-ion configurations to be the number of resolved impurity satellites, fixing the ratios of the satellite and bulk line intensities according to Eq.~(\ref{eq:P(x)}), and varying the shifts and widths of all lines for best fit. 

\begin{figure}%
%[p] \epsfxsize 6.5in \epsfbox{Liufig7.eps}			% for preprint version
[t] \epsfxsize \columnwidth \epsfbox{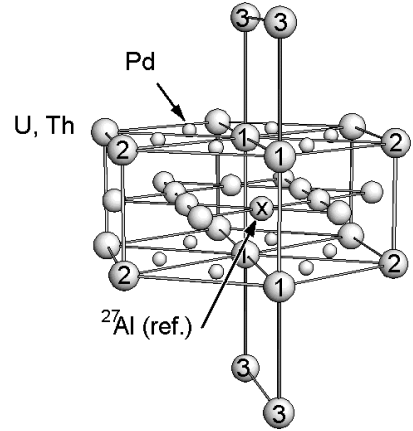}		% for two-column version
\caption{Crystal structure of U$_{1-x}$Th$_x$Pd$_2$Al$_3$. The numbers on the (U,Th) sites indicate the nearest, next-nearest, and third nearest shells to the reference $^{27}\!$Al site ({\sf X}). The latter is at the center of the rectangle formed by the four nearest-neighbor (U,Th) sites.}
\label{fig:UPd2Al3}
\end{figure}

The crystal structure of U$_{1-x}$Th$_x$Pd$_2$Al$_3$, including near-neighbor (Th,U) shells around a reference $^{27}\!$Al site out to the third shell, is shown in Figure~\ref{fig:UPd2Al3}. 
Each of the first three near-neighbor shells contains four sites ($n_0 = 4$). We found, however, that if we took $n_0 = 4$ for both the nearest-neighbor and next-nearest-neighbor shells we could not fit the spectra well. Among various possibilities we find that the choice of $n_0^{\rm nn} = 4$ for the nearest-neighbor shell and $n_0^{\rm nnn} = 8$ for the next-nearest-neighbor shell gives the best fit to the spectra. This implies that an effective next-nearest-neighbor shell is a combination of the second and third shells in the crystal structure. The distance from the reference Al site to the nearest shell is 3.402 \AA, to the second shell is 5.096 \AA, and to the third shell is 6.828 \AA, so that the difference in (U,Th)-Al distances between the second and third shells is about 1.7 \AA. This is not small on the atomic scale, and we do not understand why the RKKY coupling constants for these shells are apparently so nearly equal. 

\begin{figure}%
%[p] \begin{center}\epsfxsize 5.5in \leavevmode \epsfbox{Liufig8.eps}\end{center}% for preprint version
[t] \epsfxsize \columnwidth \epsfbox{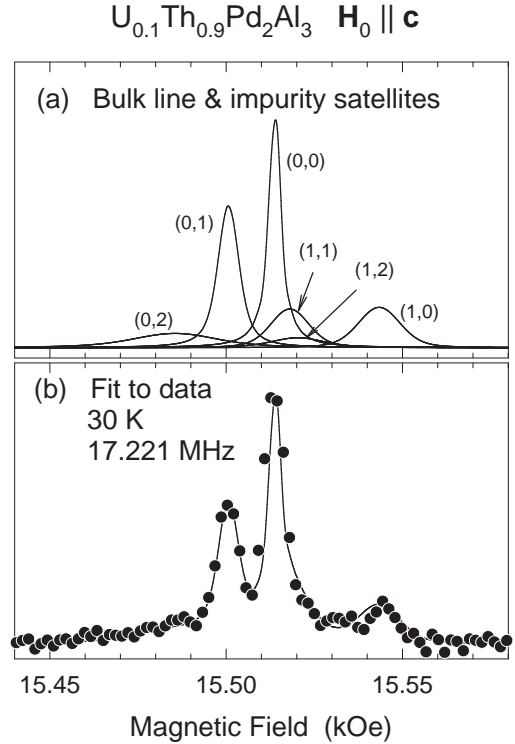}		% for two-column version
\caption{$^{27}\!$Al NMR spectrum from a field-aligned powder sample of U$_{0.1}$Th$_{0.9}$Pd$_2$Al$_3$, ${\bf H}_0\parallel {\bf c}$. (a) Bulk line and five impurity satellites included to give best fit (see text and Table~\protect\ref{tbl:P(x)} for nomenclature). (b) Circles: NMR data. Solid line: Fit curve.}
\label{fig:parrfit}
\end{figure}

For these choices of near-neighbor shell sizes, and assuming an isotropic RKKY interaction,\cite{isoRKKY} we have 5 distinct nearest-neighbor-shell configurations ($n^{\rm nn} = 0,1,2,3,4$) and 9 distinct next-nearest-neighbor-shell configurations ($n^{\rm nnn} = 0,1,2,\ldots,8$). The total number of distinct configurations associated with these two shells is therefore $5 \times 9 = 45$. Table~\ref{tbl:P(x)} gives the probabilities of the six most probable of these 45 configurations from Eq.~(\ref{eq:P(x)}) for two uranium concentrations $y=0.1$ and 0.2 ($x = 0.9$ and 0.8, respectively). 
\begin{table}[t] 
%\newpage \begin{center}\begin{minipage}{\columnwidth}
\begin{tabular}{ccr}
Configuration & $x = 0.9$ & $x = 0.8$ \\ 
\hline
$(0,0)$ & 0.2825 & 0.0687 \\ 
$(0,1)$ & 0.2510 & 0.1374 \\ 
$(1,0)$ & 0.1255 & 0.0687 \\ 
$(1,1)$ & 0.1116 & 0.1374 \\ 
$(0,2)$ & 0.0976 & 0.1203 \\ 
$(1,2)$ & 0.0434 & 0.1203 \\ 
\hline
Total & 0.9116 & 0.6528 \\
\end{tabular}
%\end{minipage}\end{center}
\caption{Probability of finding the configuration $(p,q)$ of U ions in the two nearest-neighbor (Th,U) shells about an Al site in U$_{1-x}$Th$_x$Pd$_2$Al$_3$ [from Eq.~(\protect\ref{eq:P(x)})]. Here $p$ and $q$ are the number of U ions in the nearest-neighbor and (effective) next-nearest-neighbor shells, respectively.}
\label{tbl:P(x)}
\end{table}
The configurations are designated by $(p,q)$, where $p$ and $q$ are the number of U ions in the nearest-neighbor and (effective) next-nearest-neighbor shells, respectively. We have taken the fit spectrum to include the lines corresponding to these six configurations (i.e., five impurity satellites and the bulk line) as shown in Fig.~\ref{fig:parrfit}(a). 
Fixing the relative areas of the lines according to Table~\ref{tbl:P(x)} leaves as free parameters the area of the bulk line and the shifts and linewidths of all the lines. 

The positions and widths of the lines in Fig.~\ref{fig:parrfit}(a) are the result of a fit to the spectrum of U$_{0.1}$Th$_{0.9}$Pd$_2$Al$_3$ for ${\bf H}_0\parallel {\bf c}$, $T = 30$ K, and a spectrometer frequency of 17.221 MHz. This fit is shown together with the data in Fig.~\ref{fig:parrfit}(b). It is important to note that the assignment of each line to a configuration is consistent with the oscillatory RKKY interaction. Consider for example the $(0,1)$ and $(1,0)$ impurity satellites, which are on opposite sides of the $(0,0)$ bulk line. Then one more uranium in the second shell should push the $(0,2)$ satellite further away from the $(0,0)$ line than the $(0,1)$ satellite. Similarly, the $(1,1)$ satellite should lie between the $(1,0)$ and $(0,1)$ satellites due to the opposite signs of the nearest- and next-nearest-shell interactions, and the $(1,2)$ satellite should lie between the $(1,0)$ and $(0,2)$ satellites. All these properties are satisfied by the fits without having been put in ``by hand.'' We argue that the success of this fitting procedure is excellent evidence that the satellites have been correctly identified.

\begin{figure}%
%[p] \begin{center}\epsfxsize 4.95in \leavevmode \epsfbox{Liufig9.eps}\end{center}% for preprint version
[t] \epsfxsize \columnwidth \epsfbox{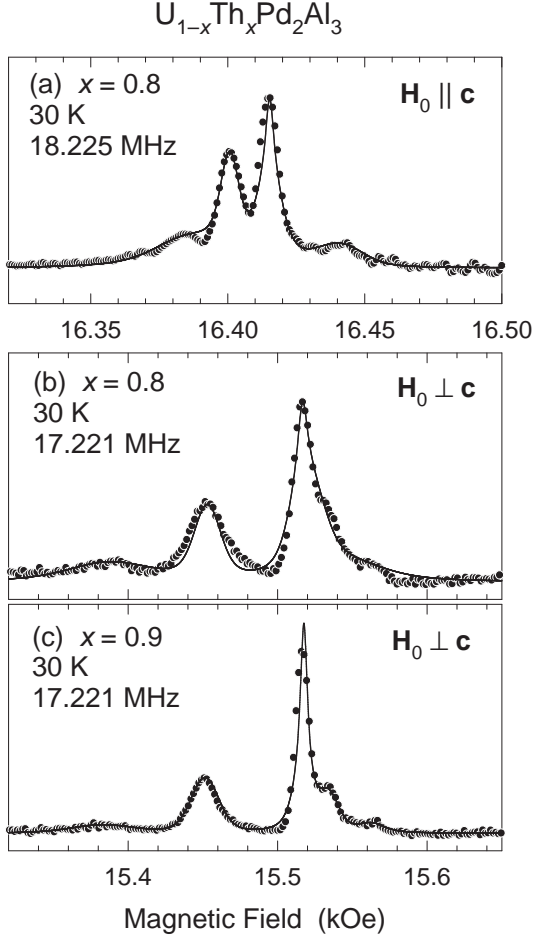}			% for two-column version
\caption{Representative fits to $^{27}\!$Al NMR field-aligned spectra from U$_{1-x}$Th$_{x}$Pd$_2$Al$_3$ alloys. (a)~$x = 0.8$, ${\bf B} \parallel {\bf c}$. (b)~$x = 0.8$, ${\bf B}\perp {\bf c}$. (c)~$x = 0.9$, ${\bf B}\perp {\bf c}$. Circles: NMR data. Solid lines: Fit curves.}
\label{fig:genfit}
\end{figure}

The model is very successful, in the sense that it gives good fits to the field-aligned spectra for both field directions and for $x = 0.9$ and $0.8$. Shown in Fig.~\ref{fig:genfit} are some examples of the spectra and their fits. 
It can be seen, however, that the fit for $x = 0.8$ is not quite as good as for $x = 0.9$. This might be associated with the fact that the total probability to find any of the six resolved-satellite U-ion neighbor configurations (cf.~Table~\ref{tbl:P(x)}) is quite high (0.9116) for $x = 0.9$, but is only 0.6528 for $x = 0.8$. The remaining 39 configurations must contribute to the broadening function, and if the probability associated with a few of these configurations is appreciable one might suspect that a simple Gaussian approximation to the line shape could break down. In line with this speculation it was found that spectra from U$_{0.3}$Th$_{0.7}$Pd$_2$Al$_3$ were even more difficult to fit. 

Even so, U$_{1-x}$Th$_x$Pd$_2$Al$_3$ is the first system in which impurity satellites have been resolved at such high magnetic impurity concentrations ($y=0.2$). This is due in part to the relatively small number of $f$-ion sites in the near-neighbor shells compared, for example, to the case in dilute {\it Cu\/}Fe alloys\cite{BoSl76}where $n_0^{\rm nn}= 12$.

\section{\boldmath Disorder and NFL behavior in U$_{1-\lowercase{x}}$T\lowercase{h}$_{\lowercase{x}}$P\lowercase{d}$_2$A\lowercase{l}$_3$} \label{sect:disorder}

Field-swept $^{27}\!$Al NMR spectra from field-aligned powder samples were obtained at a spectrometer frequency of 17.221 MHz over the temperature range~5--250 K\@. For ${\bf H}_0 \parallel {\bf c}$ all impurity satellites have very weak or temperature-independent shifts (not shown). Such behavior is expected because the $c$ axis is the magnetic ``hard'' axis.

\begin{figure}%
%[p] \epsfxsize 6.5in \epsfbox{Liufig10.eps}			% for preprint version
[t] \epsfxsize \columnwidth \epsfbox{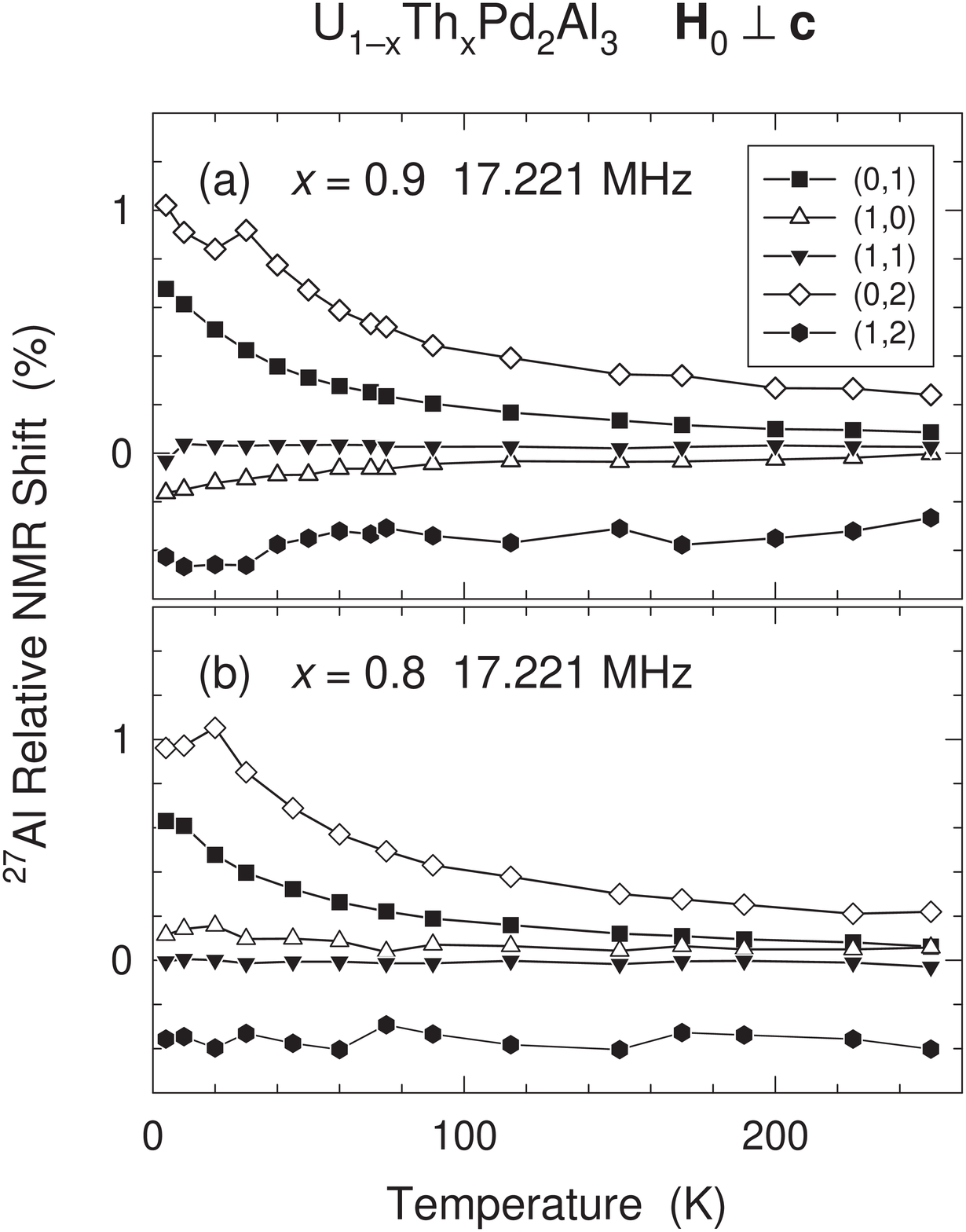}		% for two-column version
\caption{Temperature dependence of impurity satellite shifts~$K_{ab} (T)$ relative to the bulk line for ${\bf H}_0 \perp {\bf c}$ in U$_{1-x}$Th$_{x}$Pd$_2$Al$_3$ at 17.221 MHz. (a) $x = 0.9$. (b) $x = 0.8$. The $(0,1)$ and $(0,2)$ satellites have strong temperature dependences.}
\label{fig:perpKS}
\end{figure}

\begin{figure}%
%[p] \epsfxsize 6.5in \epsfbox{Liufig11.eps}			% for preprint version
[t] \epsfxsize \columnwidth \epsfbox{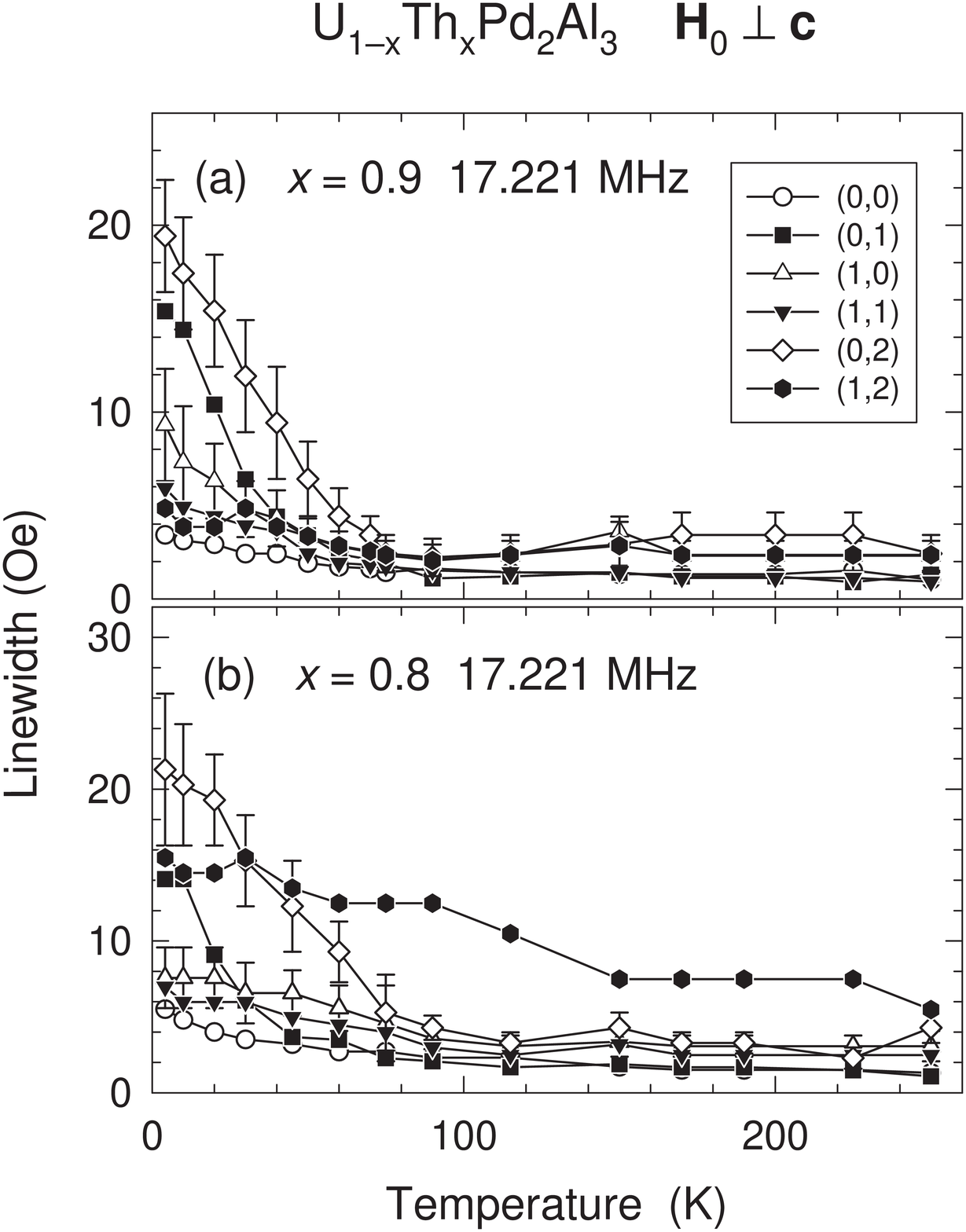}		% for two-column version
\caption{Bulk line and impurity satellite linewidths~$\sigma_{ab} (T)$ vs.~temperature for ${\bf H}_0 \perp {\bf c}$ in U$_{1-x}$Th$_{x}$Pd$_2$Al$_3$ at 17.221 MHz. (a) $x = 0.9$. (b) $x = 0.8$. The $(0,1)$ and $(0,2)$ satellite linewidths have strong temperature dependences at low $T$. Each satellite has almost the same width at high $T$ with the exception of the $(1,2)$ satellite for $x = 0.8$.}
\label{fig:perplw}
\end{figure}

Figure \ref{fig:perpKS} gives the temperature dependence of $^{27}\!$Al impurity satellite shifts~$K_{ab} (T)$ relative to the bulk line for ${\bf H}_0\perp {\bf c}$. 
There are strong similarities between results for the two concentrations, and the $(0,1)$ and $(0,2)$ satellites have stronger temperature dependences than the rest of the impurity satellites. The linewidths~$\sigma_{ab} (T)$ of impurity satellites $(0,1)$ and $(0,2)$ also have strong temperature dependence at low temperatures as shown in Fig.~\ref{fig:perplw}. 
It is not surprising to see that each satellite has almost the same width at high temperatures where the magnetic susceptibility is small. The $(1,2)$ satellite for U$_{0.2}$Th$_{0.8}$Pd$_2$Al$_3$ is an exception, as it shows only a weak temperature dependence and a larger width at high temperatures compared to the other satellites. This behavior is not understood, but it can be seen from Fig.~\ref{fig:parrfit}(a) that the $(1,2)$ satellite is the weakest of the impurity satellites and is not well resolved; there may be considerable systematic error in the parameters for this satellite.

Only the $(0,1)$ and $(0,2)$ impurity satellites have strong temperature-dependent linewidths and shifts for ${\bf H}_0\perp {\bf c}$. We use the $(0,1)$ satellite in U$_{0.1}$Th$_{0.9}$Pd$_2$Al$_3$ for further analysis, because for this satellite there is only one U moment in the immediate $^{27}\!$Al environment. This is in contrast to the situation in ligand-disorder NFL systems, e.g., UCu$_{5-x}$Pd$_x$ (Refs.~\onlinecite{MBL96}and \onlinecite{BMAF96}) and CeRhRuSi$_2$ (Refs.~\onlinecite{GTHM97}and \onlinecite{LMCN00}), where the $f$ ions are concentrated and nuclear spins couple strongly to substantially more than one neighboring $f$-ion moment. In the latter case it can be shown that the correlation length~$\xi_\chi(T)$ that characterizes the random spatial variation of the susceptibility has to be taken into account.\cite{MBL96} The use of a single-U-ion impurity satellite means that no information can be obtained concerning $\xi_\chi(T)$, but by the same token the quantity~$\sigma_{ab} (T)/K_{ab} (T)H = \delta K_{ab} (T)/K_{ab} (T)$ gives $\delta\chi(T)/\chi(T)$ independent of the (unknown) value of $\xi_\chi(T)$.

\begin{figure}%
%[p] \epsfxsize 6.5in \epsfbox{Liufig12.eps}			% for preprint version
[t] \epsfxsize \columnwidth \epsfbox{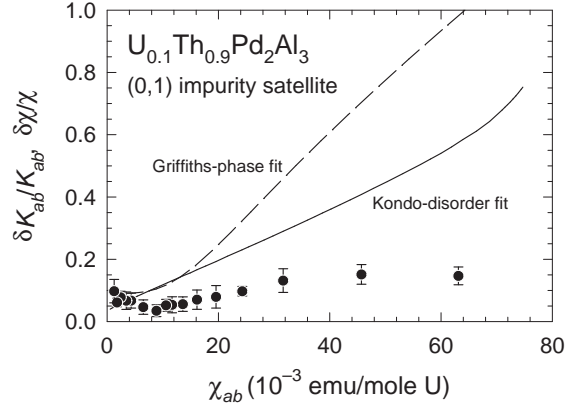}		% for two-column version
\caption{Comparison of fractional rms susceptibility spread~$\delta\chi(T)/\chi(T)$ from the Kondo-disorder (solid curve) and Griffiths-phase (dashed curve) theories with fractional $^{27}\!\!$Al NMR width~$\delta K_{ab}/K{ab}$ of the $(0,1)$ impurity satellite in U$_{0.1}$Th$_{0.9}$Pd$_2$Al$_3$, ${\bf H}_0\perp {\bf c}$ (data points). The predictions clearly overestimate the NMR data, indicating that disorder is not the major source of NFL behavior in this alloy.}
\label{fig:KDplot}
\end{figure}

Figure~\ref{fig:KDplot} plots $\delta K_{ab} (T)/K_{ab} (T)$ for the $(0,1)$ impurity satellite versus the bulk susceptibility~$\chi(T)$ in U$_{0.1}$Th$_{0.9}$Pd$_2$Al$_3$, ${\bf H}_0\perp {\bf c}$, again with temperature an implicit parameter. 
Also shown are the theoretical predictions of $\delta\chi(T)/\chi(T)$ from the single-ion Kondo-disorder and Griffiths-phase theories, obtained as described in Sec.~\ref{sect:disorderNFL}. In spite of the fact that at low temperatures the satellites are significantly broader than the central transition (Fig.~\ref{fig:perplw}), the theoretical predictions considerably overestimate the experimental results. This suggests that the effect of disorder in this system is too weak to account for its NFL behavior. The satellites are simply too narrow, a fact which, ironically, is essential to their experimental observation. Impurity satellites have also been observed in Y$_{0.8}$Th$_{0.2-y}$U$_y$Pd$_3$,\cite{LCMH98} but in this case the satellite widths become very wide and unresolved at low temperatures. We also note that in U$_{0.1}$Th$_{0.9}$Pd$_2$Al$_3$ $\delta K_{ab} (T)/K_{ab} (T)$ varies relatively slowly with $\chi(T)$ (Fig.~\ref{fig:KDplot}), i.e., the NMR linewidth is nearly proportional to the shift. This is in contrast to the behavior expected from both the Kondo-disorder and Griffiths-phase pictures, where the spread in susceptibilities is a rapidly-growing fraction of the average susceptibility as the temperature is lowered.\cite{BMLA95,CNCJ98}

\section{Conclusions}

$^{27}\!$Al NMR in the U$_{1-x}$Th$_x$Pd$_2$Al$_3$ alloy system has revealed satellite NMR lines due to specific uranium configurations around Al sites. These impurity satellites facilitate determination of the effect of disorder on paramagnetism in this system. The probability of finding a given configuration, which is related to the intensity of the corresponding line, follows a simple statistical calculation. We used a procedure that fixed the intensities according to their probabilities to fit the field-aligned spectra for $x = 0.9$ and $0.8$ and both field directions (${\bf H}_0 \parallel {\bf c}$ and ${\bf H}_0 \perp {\bf c}$). Each impurity satellite is thereby associated with a specific near-neighbor uranium configuration.

The linewidths and shifts in U$_{1-x}$Th$_x$Pd$_2$Al$_3$ do not have much temperature dependence for ${\bf H}_0\parallel {\bf c}$, as is also the case for the $c$-axis susceptibility~$\chi_c(T)$. In contrast, two of the impurity satellites have Curie-Weiss-like temperature-dependent linewidths and shifts for ${\bf H}_0\perp {\bf c}$. But the linewidths do not increase much more rapidly with decreasing temperature than the shift, so that $\delta K/K$ does not exhibit the rapid increase with bulk susceptibility $\chi$ expected from the disorder-driven models.\cite{MBL96,CNCJ98} These mechanisms also overestimate the observed linewidth at low temperatures, suggesting that the disorder in this system is not strong enough to account for its NFL behavior. 

In the disorder-driven models the origin of the disorder is variation of the $f$-electron/conduction-electron hybridization matrix element with local $f$-ion environment. Disorder is found to be an important contributor to NFL behavior in alloys with ligand disorder, such as UCu$_{5-x}$Pd$_x$ (Ref.~\onlinecite{MBL96}) and CeRhRuSi$_2$ (Ref.~\onlinecite{LMCN00}). One might suspect, however, that the immediate $f$-ion environment is not as strongly disordered in dilute solid solutions of $f$ ions, given that in the limit of infinite dilution all $f$ ions have identical environments. NFL behavior in such systems might therefore be due to some other mechanism. This conclusion must be regarded as speculative, however, since U concentrations of 10\% and 20\% can hardly be considered dilute. In addition, U$_{1-x}$Th$_x$Pd$_2$Al$_3$ is the only $f$-ion diluted system studied to date using NMR.

NFL behavior in U$_{1-x}$Th$_x$Pd$_2$Al$_3$ is observed over a wide range of Th concentrations, and the thermodynamic and transport properties obey single-ion scaling.\cite{MdAHD95} These results suggest that a quantum critical point associated with cooperative behavior is not the NFL mechanism in this system. Two models that rely neither on cooperative effects nor on disorder have been applied specifically to U$_{1-x}$Th$_x$Pd$_2$Al$_3$. These are (a)~the quadrupolar Kondo model,\cite{MdAHD95} which assumes a non-Kramers doubly degenerate nonmagnetic ground state, and (b)~the electronic polaron model of Liu,\cite{Liu97} which assumes that the $f$-electron energies are close to the Fermi level and that transport involves polaron-like hopping between $f$ sites. Neither of these theories predicts strong disorder in the magnetic susceptibility, and from the standpoint of our NMR results both remain candidates for NFL behavior in U$_{1-x}$Th$_x$Pd$_2$Al$_3$.

\section*{Acknowledgments}
We are grateful to A.~H. Castro Neto for helpful discussions and comments. This research was supported by the U.S.~NSF, Grant nos.~DMR-9418991 (U.C. Riverside) and DMR-9705454 (U.C. San Diego), by the U.C. Riverside Academic Senate Committee on Research, and by the Research Corporation (Whittier College). 

%\bibliographystyle{c:/PCTeXv4/TeXBIB/prsty}				% PR/PRL style.
%\bibliography{c:/PCTeXv4/TeXBIB/UTPd2Al3,c:/PCTeXv4/TeXBIB/NFL,c:/PCTeXv4/TeXBIB/muSR,c:/PCTeXv4/TeXBIB/NMR,c:/PCTeXv4/TeXBIB/CeRR2Si2,c:/PCTeXv4/TeXBIB/KondoEff}

\begin{thebibliography}{10}
\bibitem{KSdef}
This NMR shift is sometimes called the Knight shift, but we prefer to reserve
 the latter term exclusively for the shift due to conduction electrons in
 metals.
\bibitem{isoRKKY}
Anisotropy in the RKKY interaction would further powder-pattern broaden the
 satellite lines proportionally to the susceptibility. We will see in
 Sec.~{\protect\ref{sect:disorder}} that any such contribution to the
 linewidth does not affect the conclusions of this paper.
\end{thebibliography}

%\end{document}

\end{document}